\documentclass[sigconf,nonacm]{acmart}

\setcopyright{none}
\renewcommand\footnotetextcopyrightpermission[1]{}
\title{Multidimensional Views Behind Target-Price Revisions:   \\An LLM-Based Representation of Analyst Outlooks}

\author{Kiyoun Kim}
\email{kiyoun.kim@kic.go.kr}
\affiliation{%
  \institution{Korea Investment Corporation}
  \country{Republic of Korea}
}

\author{Sung-Yeon Nathan Lim}
\email{nathan.lim@kic.go.kr}
\affiliation{%
  \institution{Korea Investment Corporation}
  \country{Republic of Korea}
}

\renewcommand{\shortauthors}{Kim and Lim}

\usepackage[skins,breakable]{tcolorbox}
\newtcolorbox{promptbox}[1]{
  enhanced,breakable,
  colback=black!2,colframe=black!35,
  colbacktitle=black!8,coltitle=black,
  fonttitle=\small\bfseries,fontupper=\footnotesize,
  title={#1},boxrule=0.4pt,arc=1.5mm,
  left=2mm,right=2mm,top=1.5mm,bottom=1.5mm,
  before skip=8pt,after skip=8pt
}

\begin{document}

\begin{abstract}
A target price compresses an analyst's multidimensional views into a single number. We propose an LLM-based text representation that extracts these views from research reports as question-based outlook scores with supporting evidence. The questions organize reported views on earnings, valuation, industry conditions, and other investment dimensions into a compact representation of analyst sentiment. In walk-forward evaluation, the scores improve prediction of subsequent target-price revisions beyond current-revision momentum and sector differences. They outperform financial dictionary and FinBERT sentiment measures. These findings suggest that structured questions can capture information about subsequent analyst actions that is not fully reflected in the current numerical revision. By linking each dimension to an explicit investment question and supporting report evidence, the approach makes the extracted outlooks available for inspection and interpretation. It offers a way to complement target-price signals with a multidimensional view of the judgments expressed in analyst narratives.
\end{abstract}


\maketitle

\section{Introduction}

Sell-side equity research reports communicate analysts' assessments of listed companies, including earnings and revenue expectations, profitability, shareholder distributions, industry conditions, and competitive positioning. A target price summarizes the analyst's valuation judgment, while the narrative explains its assumptions and risks. These written assessments contain information beyond numerical forecasts and recommendations \cite{asquith2005information}. A single target price cannot fully represent potentially conflicting views: stronger revenue expectations may coexist with weaker margins or less favorable valuation assumptions. Examining the narrative can therefore reveal distinct dimensions of the analyst's outlook. These dimensions may change even when the numerical target remains unchanged, making the text a potential source of information about later valuation updates.

\begin{figure*}[!t]
\centering
\includegraphics[width=\textwidth]{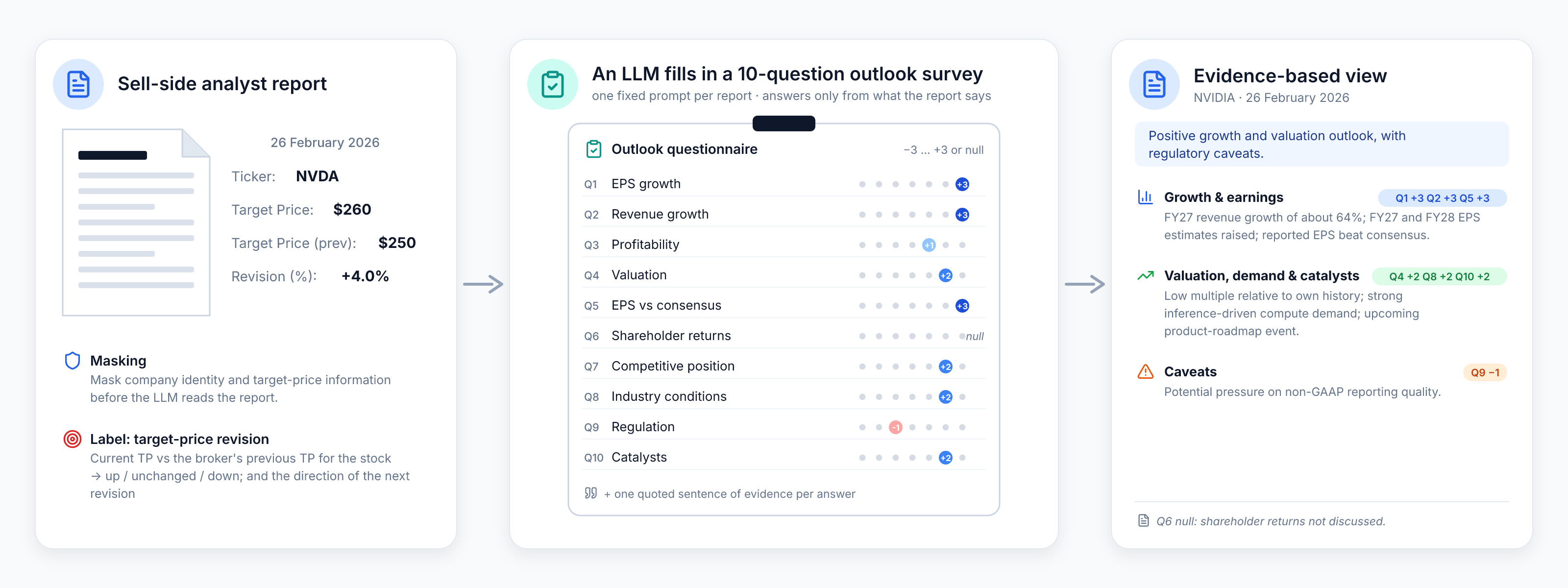}
\caption{Analysis example from the NVIDIA research report dated 26 February 2026.}
\label{fig:nvda-workflow-examples}
\Description{The NVIDIA report dated 26 February 2026 illustrates company and target-price masking, ten question-based outlook scores, and an evidence-based interpretation accompanying an upward target-price revision.}
\end{figure*}

In quantitative investment practice, target-price revisions provide signals of changing analyst views. For a given brokerage--company series, we define the revision between consecutive reports as

\begin{equation}
\mathrm{TPRev}_{t} = \frac{\mathrm{TP}_{t}-\mathrm{TP}_{t-1}}{\mathrm{TP}_{t-1}},
\end{equation}

where $\mathrm{TP}_{t}$ and $\mathrm{TP}_{t-1}$ are the current and preceding target prices. Positive and negative revisions indicate raises and cuts. Practitioner research examines revision-based signals, including changes in the target-to-market-price gap, for return prediction \cite{tortoriello2019bridges}. However, analyst outputs can reflect herding in earnings forecasts and consensus effects in recommendations \cite{hong2000career,welch2000herding}. Even an independently formed revision compresses multiple judgments into one value, obscuring which aspects of the outlook have changed.

Analyst questionnaires offer a way to separate these judgments. Such questionnaires collect responses directly from analysts on industry and regulatory conditions and short-term catalysts, assigning scores by company and question \cite{ubsQuantResearchTearsheets}. Extending such surveys requires analyst participation and administrative effort. We instead investigate whether an LLM can use published reports to produce similarly structured outlook indicators. A fixed questionnaire provides a common basis for comparing assessments across reports while preserving differences among investment dimensions.

We propose an LLM-based approach that answers ten investment-oriented questions, refined through expert review, using each report as evidence. The resulting question-based outlook scores (QOS) represent distinct investment judgments, with a supporting statement for every score. We evaluate their relationship with current target-price actions and, primarily, their ability to predict subsequent revisions beyond current revision and sector controls. Comparisons with sentiment measures, text embeddings, and alternative masking conditions assess the information retained by this compact representation. The evidence statements make each assessment available for inspection. Our objective is to evaluate whether this interpretable representation captures information relevant to subsequent analyst actions. Figure~\ref{fig:nvda-workflow-examples} illustrates the workflow using an NVIDIA report.

Section 2 reviews related work; Sections 3 and 4 describe the data and extraction framework. Section 5 presents the evaluation and results, Section 6 discusses applications, and Sections 7 and 8 address limitations and conclusions.

\section{Related Work}

Prior research establishes that analyst narratives convey information beyond numerical forecasts and recommendations. Huang, Zang, and Zheng (2014) \cite{huang2014text} use a naive Bayes classifier to measure textual opinions in analyst reports and find that the text provides incremental information beyond contemporaneous earnings forecasts, recommendations, and target prices, while also helping investors interpret these signals. Huang, Lehavy, Zang, and Zheng (2018) \cite{huang2018roles} use topic modeling to compare analyst reports with earnings conference calls. They document both information discovery, through discussion of topics beyond corporate disclosures, and interpretation of information already disclosed. Together, these studies motivate examining the distinct judgments embedded in report narratives. Our approach represents those judgments along dimensions defined by investment-oriented questions.

Recent work applies LLMs to the investment information in analyst reports and examines their sensitivity to contextual opinions. Lv (2025) \cite{lv2025investment} combines LLM embeddings with machine-learning models to forecast future long-term stock returns. The study finds information incremental to analysts' numerical outputs and established predictors, and uses a Shapley decomposition to examine topic contributions, highlighting strategic outlook and forward-looking fundamental assessments. Hu and Zhao (2026) \cite{hu2026finbias} introduce Fin-Bias and evaluate LLM investment judgments using analyst reports with original ratings, ratings removed, or manipulated ratings. Their findings show that explicit analyst opinions can influence model-generated ratings. These studies motivate both the use of LLM-based text representations and careful assessment of their dependence on headline opinions. Our representation assigns an explicit meaning to each score through a predefined question and links the score to supporting report evidence.

Xiao and Zang (2025) \cite{xiao2025vague} provide the closest precedent for predicting subsequent analyst revisions. In Vague Knowledge: Evidence from Analyst Reports, they argue that information that is difficult to express precisely may appear in an analyst's language before entering numerical forecasts. Their analysis relates current report tone to subsequent earnings-forecast, target-price, and recommendation revisions by the same analyst, controlling for contemporaneous revisions and other characteristics. In particular, they document a positive association between current tone and subsequent target-price revisions.

Building on this evidence, we examine whether a ten-dimensional, question-based representation adds predictive information beyond the current target-price revision in a walk-forward evaluation. Our focus is on measuring and evaluating structured outlooks, with textual evidence enabling inspection of the extracted scores.

\section{Data}

\subsection{Research Reports and Target Prices}

The sample comprises 41,640 single-stock equity research reports on US-listed companies, issued by one global brokerage between January 2017 and August 2026 and covering 1,807 stocks. We include reports with an observed subsequent target price for the same stock and exclude current target-price revisions of 50\% or more in absolute magnitude. Coverage spans a median of 3.5 years, measured from the first to the last report for each stock in the sample, with a median of 15 reports per stock. Each report has an associated publication date and investment rating, and each stock is assigned a Global Industry Classification Standard (GICS) sector.

For each report, we calculate the current target-price revision from the published target price and the associated prior target price. We retain reports whose prior target price was published within the preceding 180 days; the median interval to that prior observation is 27 days. We treat reports on the same stock as a single series even when the report authors change. In the final sample, 62.6\% of current target prices are unchanged, 22.6\% are raised, and 14.8\% are lowered. For the next observed revision, the corresponding proportions are 55.2\%, 25.6\%, and 19.3\%. This prevalence of unchanged targets motivates evaluating upward and downward revisions separately against an unchanged reference category.

\begin{figure*}[t]
\centering
\includegraphics[width=\textwidth]{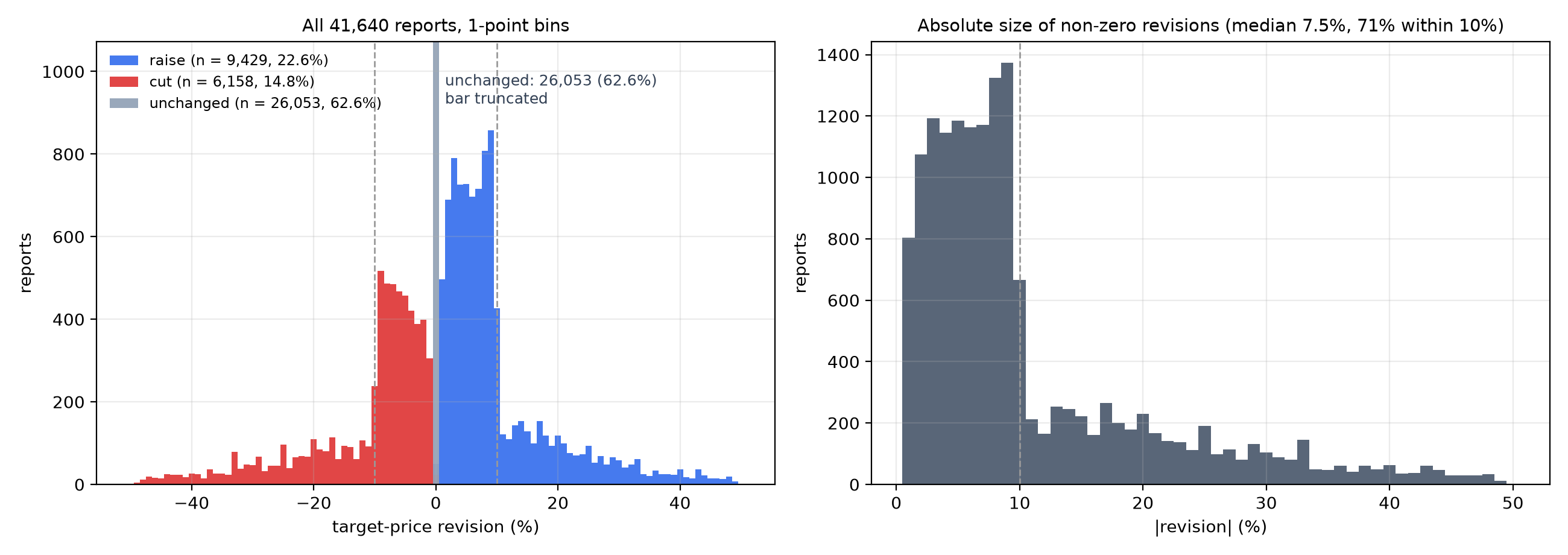}
\caption{Distribution of current target-price revisions in the final sample of 41,640 reports. Left: revisions in one-percentage-point bins; the unchanged-target bar is truncated to show the distribution of non-zero revisions. Right: absolute magnitudes of non-zero revisions. Dashed lines mark 10\% magnitudes.}
\label{fig:rev-distribution}
\Description{Histograms of signed and absolute target-price revisions, with the unchanged-target bar truncated. Non-zero revisions have a median absolute change of 7.5 percent and 71 percent are within 10 percent.}
\end{figure*}

Figure~\ref{fig:rev-distribution} shows the distribution of current target-price revisions. Among non-zero revisions, the median absolute change is 7.5\%, and 71\% are within 10\%.

Table~\ref{tab:1} summarizes the transition probabilities from the current target-price revision to the next observed revision. Unchanged targets remain the most likely next outcome following every current revision direction, with probabilities ranging from 49.8\% to 57.9\%. Nevertheless, an upward revision is followed by another upward revision in 37.4\% of cases, compared with 23.4\% following an unchanged target. Similarly, the probability of a subsequent downward revision is 31.6\% following a downward revision, compared with 18.7\% following an unchanged target. These transition patterns indicate directional persistence, or momentum, in the brokerage's target-price actions and motivate controlling for the current revision when evaluating the additional information in report text.

\begin{table}[!ht]
\caption{Transition probabilities from the current target-price revision to the next observed revision. Row percentages for the final sample of 41,640 reports; $N$ denotes the number of reports.}
\label{tab:1}
\centering
\small
\setlength{\tabcolsep}{4pt}
\begin{tabular}{@{}lrrrr@{}}
\toprule
& \multicolumn{3}{c}{\textbf{Next revision}} & \\
\cmidrule(lr){2-4}
\textbf{Current revision} & \textbf{Up} & \textbf{Unchanged} & \textbf{Down} & $N$ \\
\midrule
Up & 37.4\% & 49.8\% & 12.8\% & 9,429 \\
Unchanged & 23.4\% & 57.9\% & 18.7\% & 26,053 \\
Down & 16.6\% & 51.7\% & 31.6\% & 6,158 \\
\midrule
All & 25.6\% & 55.2\% & 19.3\% & 41,640 \\
\bottomrule
\end{tabular}
\end{table}

\subsection{Text Preparation}

We extract text from the report PDFs and retain the title, summary, and body paragraphs. We exclude tables, figures, and material unrelated to the report's analysis, such as disclosures and disclaimers. Short fragments are removed. This procedure emphasizes narrative assessments while excluding much of the standardized material surrounding them.

\section{Methodology}

\subsection{Question-Based Outlook Extraction}

The framework converts each report into a vector of ten question-based outlook scores (QOS). The questions cover financial performance, valuation, shareholder distributions, competitive positioning, industry conditions, regulation, and catalysts. Each question defines the assessment to be extracted, such as expected year-on-year EPS growth or earnings relative to market consensus, so that scores reflect specific analyst views rather than the report's overall tone. Table 2 summarizes the dimensions and their interpretation.

\begin{table*}[t]
\caption{Questionnaire dimensions. Positive scores indicate the favorable assessment described below; negative scores indicate its opposite.}
\label{tab:2}
\centering
\small
\begin{tabular}{@{}p{0.08\textwidth}p{0.19\textwidth}p{0.64\textwidth}@{}}
\toprule
\textbf{Question} & \textbf{Dimension} & \textbf{Basis for a positive assessment} \\
\midrule
Q1 & EPS growth & Expected increase relative to the comparable prior-year period \\
Q2 & Revenue growth & Expected increase relative to the comparable prior-year period \\
Q3 & Profitability & Expected improvement in returns on capital or margins relative to the prior-year period \\
Q4 & Valuation & Current share price assessed as undervalued \\
Q5 & Earnings versus consensus & Next reported EPS expected to exceed current market consensus \\
Q6 & Shareholder returns & Dividends or repurchases expected to increase relative to the prior-year period \\
Q7 & Competitive position & Expected strengthening of the company's position relative to peers \\
Q8 & Industry environment & Expected improvement in industry demand, pricing, competition, capacity, or input costs \\
Q9 & Government and regulation & Expected improvement in policy or regulatory conditions \\
Q10 & Catalysts & Identified future developments expected to have a positive share-price impact \\
\bottomrule
\end{tabular}
\end{table*}

For report $t$ and question $j$, the model returns a score $s_{t,j}$ on a seven-point scale from $-3$ to $+3$, together with a short quotation or paraphrase supporting that score. The endpoints represent strongly unfavorable and strongly favorable assessments. A score of zero is reserved for an explicitly neutral, unchanged, in-line, or balanced view. Missing or unclear evidence receives a null value. This distinction prevents the extraction step from treating the absence of discussion as an expressed neutral opinion. Missingness varies substantially across dimensions: approximately 65\% of reports do not discuss shareholder returns, compared with 3\% for catalysts.

We also considered alternative treatments for null scores, including question-specific missingness indicators and carrying forward previously observed scores, but these yielded similar predictive performance to replacing null scores with zero. We therefore use zero replacement in this study.

The model is instructed to use only the supplied report, follow the comparison basis of each question, and extract the analyst's stated or clearly implied assessment. All ten questions are answered in a single call using Qwen3.5-122B-A10B with temperature set to zero and reasoning mode disabled. The output contains the ten scores and their supporting evidence statements. The evidence is retained for inspection; the statistical models use the scores alone. Appendix A reproduces the scoring prompt and the operative masking instructions.

\begin{figure*}[t]
\centering
\includegraphics[width=\textwidth]{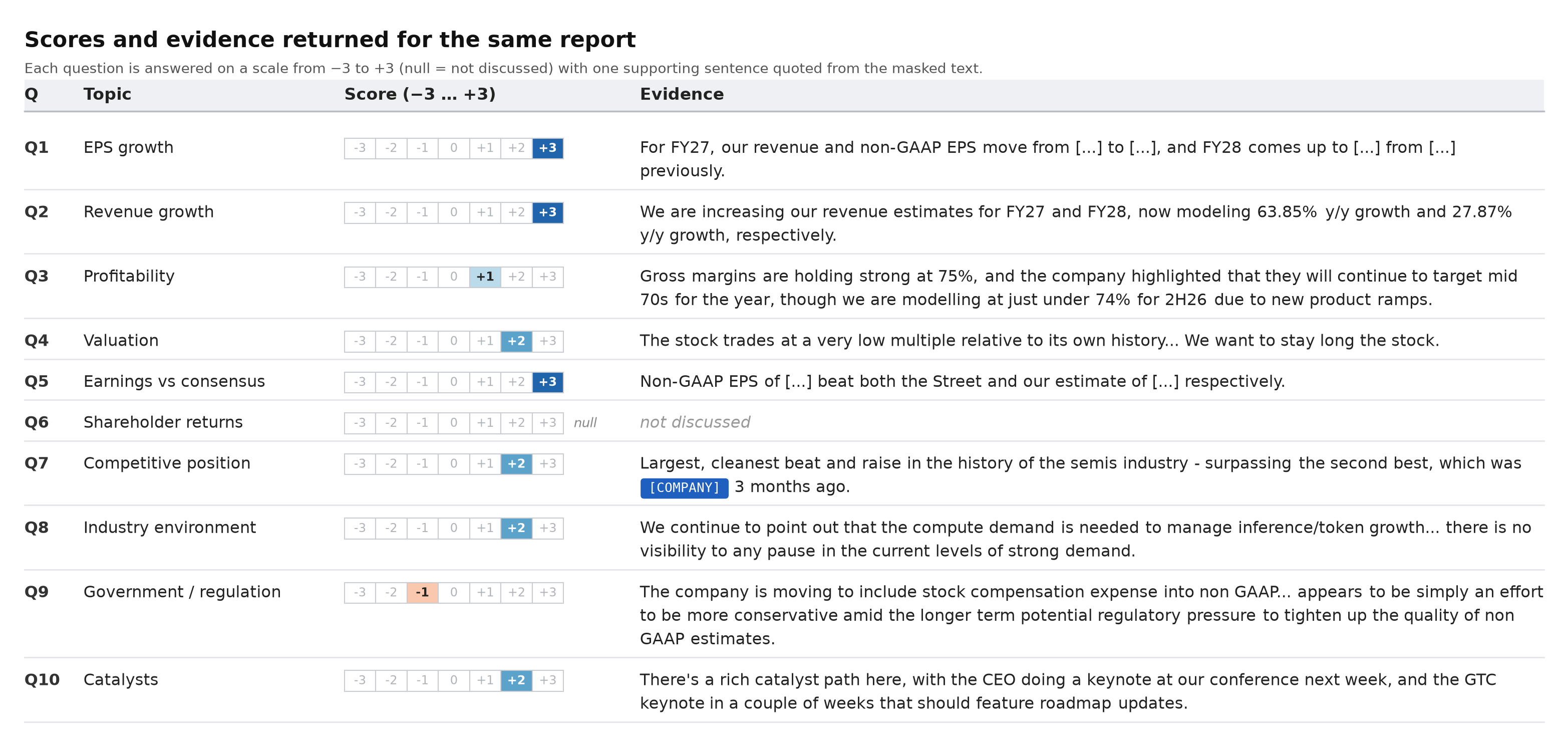}
\caption{Model-generated outlook scores and supporting evidence from the masked report. Highlighted positions show the seven-point assessments; null denotes missing or unclear evidence. Selected monetary amounts, per-share values, and valuation multiples in the displayed evidence are omitted as [...].}
\label{fig:scoring-example}
\Description{Example of question-based scores and supporting evidence}
\end{figure*}

Figure~\ref{fig:scoring-example} illustrates the extracted scores and evidence for the NVIDIA report dated 26 February 2026. Each question produces a separate assessment, while topics without supporting discussion receive null values. The associated evidence makes the textual basis of each model-generated score available for review.

\subsection{Masking Company Identifiers and Target-Price Information}

Explicit statements of target-price actions can create a mechanical relationship between the report text and the contemporaneous outcome. Before scoring, we therefore mask the covered company's name and ticker, target-price amounts, expressions announcing target-price increases, decreases, or maintenance, and target-price-derived upside or downside percentages. A language model identifies the relevant text spans, which are checked against the original text and replaced with category-specific placeholders. Other financial forecasts, valuation multiples, and narrative explanations are retained.

The primary specification uses both company and target-price masking. To examine sensitivity, we also score reports under three alternative conditions: no masking, company masking only, and target-price masking only. Company and target-price masking are also intended to reduce reliance on company-specific knowledge and knowledge of target-price actions acquired during LLM pretraining. The procedure is designed to limit reliance on information learned after a report's publication that could introduce look-ahead bias, as well as direct access to the numerical opinion being evaluated.

A side-by-side masking example is provided in Appendix A (Figure~\ref{fig:masking-example}).

\section{Experiments and Results}

\subsection{Evaluation Design}

The empirical outcomes are target-price revisions rather than stock returns. We therefore focus on two tasks: identifying the contemporaneous target-price revision and predicting the subsequent revision. The first tests whether the scores capture views associated with the current target-price decision; the second examines whether the narrative contains information about later target-price actions. Predictive information about subsequent revisions would be consistent with emerging valuation judgments or intentions being reflected in the text before a numerical update.

We use L2-regularized logistic regression as the primary classifier. For contemporaneous classification, regularization strengths are fixed, with stronger regularization for the higher-dimensional embeddings; performance is evaluated using five-fold cross-validation grouped by stock, so that reports on a given stock appear in only one held-out fold.

For subsequent-revision prediction, we select the regularization strength using a purged, time-ordered validation set. Each rolling 36-month training window is divided chronologically into an initial 24-month fitting period and a final 12-month validation period. Fitting observations are retained only if their subsequent target-price outcome is observed before the validation period begins; validation outcomes must be observed before the test quarter begins. We select the value with the highest validation AUC, refit on the full 36-month window using all outcomes observed before the test quarter, and evaluate on that quarter.

For subsequent revisions, the current revision is entered as a continuous percentage change alongside GICS sector indicators. We use current revision alone and current revision with GICS sector indicators as simple baselines to capture revision momentum and sector differences.

Four text representations serve as baselines: Loughran--McDonald financial dictionary measures \cite{loughran2011liability}, FinBERT sentiment \cite{araci2019finbert}, BGE-M3 embeddings \cite{chen2024m3}, and FinLang/finance-embeddings-investopedia (hereafter FinLang) \cite{finlangInvestopedia}. The dictionary representation comprises net tone and positive, negative, and uncertainty word shares. FinBERT produces sentiment probabilities aggregated across text chunks. BGE-M3 and FinLang produce 1,024- and 768-dimensional embeddings, respectively, and are evaluated with regularized logistic regression. All baselines are computed from the same retained, fully masked report text, excluding formatting markers and masking placeholders, subject to each encoder's input limits.

We assess classification performance using the area under the receiver operating characteristic curve (AUC). For contemporaneous classification, AUC is computed from pooled held-out predictions; for subsequent-revision prediction, we report the mean quarterly AUC.

\subsection{Contemporaneous Target-Price Revisions}

Table~\ref{tab:3} compares contemporaneous revision classification across three binary tasks. The experiment measures how well the extracted outlooks distinguish the target-price action reported in the same document.

\begin{table}[t]
\caption{Contemporaneous target-price classification (AUC). U, D, and N denote upward, downward, and unchanged revisions; TP denotes target price and LM denotes Loughran--McDonald.}
\label{tab:3}
\centering
\small
\setlength{\tabcolsep}{3pt}
\begin{tabular*}{\columnwidth}{@{\extracolsep{\fill}}lrrr@{}}
\toprule
\textbf{Representation} & \textbf{U / D} & \textbf{U / N} & \textbf{D / N} \\
\midrule
QOS, no mask & 0.860 & 0.716 & 0.737 \\
QOS, company mask & 0.860 & 0.716 & 0.736 \\
QOS, TP mask & 0.860 & 0.719 & 0.734 \\
QOS, full mask & 0.859 & 0.718 & 0.734 \\
LM dictionary & 0.740 & 0.647 & 0.599 \\
FinBERT & 0.773 & 0.586 & 0.671 \\
BGE-M3 & 0.869 & 0.876 & 0.890 \\
FinLang & 0.845 & 0.844 & 0.859 \\
\bottomrule
\end{tabular*}
\end{table}

QOS outperforms the dictionary and FinBERT baselines on all three tasks. Its discrimination between upward and downward revisions is comparable to the embedding models, despite using only ten interpretable scores. Embeddings perform substantially better when distinguishing either revision direction from unchanged targets.

Performance is similar across the four masking conditions, suggesting that aggregate discrimination is not strongly dependent on the explicit company and target-price information removed here. This does not imply that individual scores are unaffected by masking. Score-level revision profiles are provided in Figure~\ref{fig:score-current-revision} in the appendix.

\subsection{Predicting Subsequent Target-Price Revisions}

The prediction experiment evaluates the direction of the next observed target-price update, separately for upward versus unchanged and downward versus unchanged outcomes. The evaluation spans 27 quarterly folds from 2020Q1 through the observed portion of 2026Q3.

Table~\ref{tab:4} reports the validation-tuned logistic regression results. The current-revision baseline achieves mean quarterly AUCs of 0.557 for upward revisions and 0.551 for downward revisions. Adding sector controls raises these values to 0.621 and 0.579. Relative to this stronger baseline, QOS increases AUC to 0.640 and 0.612, gains of 0.019 and 0.033, respectively. Thus, the gain over current revision alone should not be attributed entirely to the text scores.

\begin{table}[t]
\caption{Subsequent target-price prediction (mean quarterly AUC). Controls denote current revision and GICS sector; Up and Down are each compared with unchanged targets.}
\label{tab:4}
\centering
\small
\setlength{\tabcolsep}{4pt}
\begin{tabular*}{\columnwidth}{@{\extracolsep{\fill}}lrr@{}}
\toprule
\textbf{Predictors} & \textbf{Up} & \textbf{Down} \\
\midrule
Current revision & 0.557 & 0.551 \\
Controls & 0.621 & 0.579 \\
Controls + Loughran--McDonald & 0.627 & 0.588 \\
Controls + FinBERT & 0.622 & 0.597 \\
Controls + QOS & 0.640 & 0.612 \\
Controls + BGE-M3 & 0.658 & 0.630 \\
Controls + FinLang & 0.656 & 0.623 \\
\bottomrule
\end{tabular*}
\end{table}

With the same current-revision and sector controls, QOS outperforms the dictionary and FinBERT representations on both tasks and provides incremental predictive information about subsequent target-price revisions. Although its AUCs are modestly lower than those of BGE-M3 and FinLang, the ten scores retain much of the predictive discrimination achieved by the higher-dimensional embeddings. QOS therefore offers a compact, interpretable representation of much of the report information relevant to predicting subsequent analyst actions.

\subsection{Prediction Performance by Sector}

\begin{figure*}[t]
\centering
\includegraphics[width=0.85\textwidth]{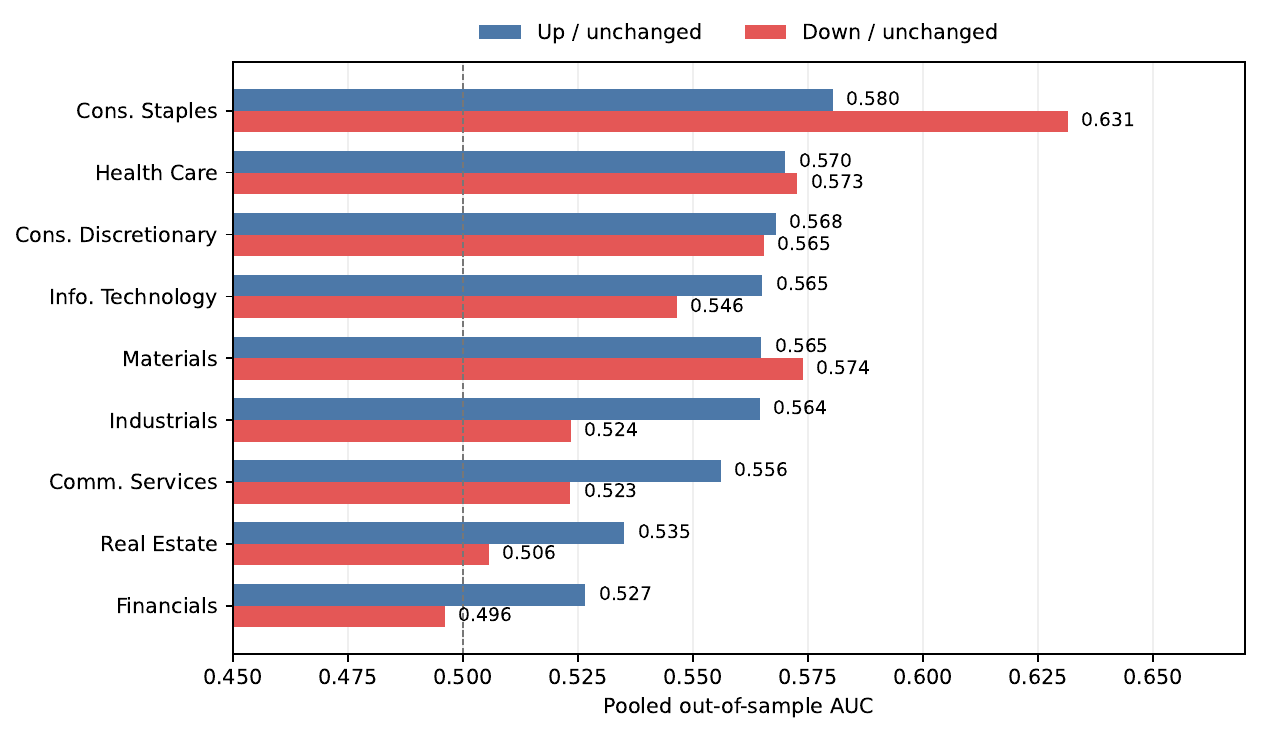}
\caption{Prediction performance of separately trained GICS-sector models using current revision and QOS. Bars show pooled out-of-sample AUCs over available quarterly test predictions.}
\label{fig:sector-refit}
\Description{Paired horizontal bars show upward and downward revision AUCs for nine separately trained sector models. Consumer Staples has the highest AUC in both tasks.}
\end{figure*}

We fit separate models within each GICS sector using current revision and QOS under the same walk-forward evaluation framework. Figure~\ref{fig:sector-refit} summarizes pooled out-of-sample AUCs within each sector.

Consumer Staples shows the strongest discrimination, with AUCs of 0.580 for upward revisions and 0.631 for downward revisions. Health Care also performs relatively well in both directions (0.570 and 0.573), while Consumer Discretionary and Information Technology are among the stronger sectors for upward revisions. Performance is weaker in Financials and Real Estate, particularly for downward revisions.

One possible explanation is the fit between the questionnaire and the investment judgments emphasized in each sector. Its focus on earnings, revenue growth, and profitability may help capture outlooks in Consumer Staples, Consumer Discretionary, and Information Technology, while the catalyst question may be relevant to event-driven assessments in Health Care. Conversely, the weaker results in Real Estate and Financials suggest considering additional sector-specific questions, such as capitalization rates and occupancy for real estate, or net interest margins, credit quality, and capital adequacy for banks.
\subsection{Explanatory Power of Individual Questions}

Table~\ref{tab:question-coefficients} summarizes the conditional associations of individual questions with current and subsequent target-price revisions. The coefficients come from separate full-sample regressions, rather than the walk-forward prediction models.

\begin{table*}[t]
\caption{Standardized question-level coefficients for current and subsequent target-price revisions.}
\label{tab:question-coefficients}
\centering
\small
\begin{tabular}{@{}lrrrr@{}}
\toprule
\textbf{Question} & \textbf{Current up} & \textbf{Current down} & \textbf{Next up} & \textbf{Next down} \\
\midrule
Q1 EPS growth & $+0.39^{**}$ & $-0.31^{**}$ & $+0.04$ & $-0.12^{**}$ \\
Q2 Revenue growth & $+0.13^{**}$ & $+0.26^{**}$ & $-0.01$ & $+0.12^{**}$ \\
Q3 Profitability & $+0.06^{*}$ & $-0.19^{**}$ & $+0.12^{**}$ & $+0.06^{**}$ \\
Q4 Valuation & $-0.33^{**}$ & $+0.31^{**}$ & $-0.27^{**}$ & $+0.03$ \\
Q5 Earnings vs consensus & $+0.43^{**}$ & $-0.31^{**}$ & $+0.16^{**}$ & $-0.15^{**}$ \\
Q6 Shareholder returns & $+0.11^{**}$ & $+0.06^{**}$ & $+0.02$ & $+0.04^{*}$ \\
Q7 Competitive position & $+0.10^{**}$ & $-0.09^{**}$ & $+0.07^{**}$ & $-0.02$ \\
Q8 Industry environment & $+0.07^{**}$ & $-0.40^{**}$ & $+0.08^{**}$ & $-0.13^{**}$ \\
Q9 Government and regulation & $+0.06^{**}$ & $-0.00$ & $-0.02$ & $+0.04^{*}$ \\
Q10 Catalysts & $-0.06^{*}$ & $-0.26^{**}$ & $+0.06^{**}$ & $-0.17^{**}$ \\
\bottomrule
\end{tabular}
\par\smallskip
\begin{minipage}{0.98\textwidth}
\footnotesize Notes: Standardized coefficients from unpenalized logistic regressions using the available full sample; null scores are set to zero. Each task compares an upward or downward revision with an unchanged target. Current-revision models include the ten scores; subsequent-revision models additionally control for current revision and GICS sector. Control coefficients are omitted. Standard errors are clustered by stock. $^{*}p<0.05$; $^{**}p<0.01$.
\end{minipage}
\end{table*}

EPS growth (Q1) and earnings relative to consensus (Q5) are positively associated with current raises, while weaker industry conditions (Q8), earnings views (Q1 and Q5), and catalysts (Q10) accompany current cuts. For subsequent revisions, the consensus assessment (Q5) remains associated with both directions. Profitability (Q3) is positively associated with subsequent raises, while unfavorable catalysts (Q10) and industry conditions (Q8) are associated with subsequent cuts. EPS growth (Q1) and revenue growth (Q2) are not statistically significant for subsequent raises in this specification.

One interpretation is that EPS growth (Q1) is already reflected in the current target-price revision, leaving limited additional information for predicting subsequent raises. Earnings relative to consensus (Q5), by contrast, captures expected beats or misses that may be confirmed at a later earnings announcement and prompt a target-price update. Negative catalysts (Q10) may similarly identify risks that appear in the narrative before their implications are incorporated into a subsequent target-price cut.

Valuation (Q4) is negatively associated with raises after controlling for the other scores, despite its generally positive unconditional relationship with current revisions. A possible explanation is that describing a stock as undervalued can support retaining an existing target without raising it.

Sector-level results suggest that the outlook dimensions associated with subsequent target-price revisions vary across industries. Industry conditions (Q8) are particularly prominent for upward revisions in Materials and downward revisions in Industrials. Catalysts (Q10) are more prominent in Health Care and consumer sectors, especially for subsequent cuts, with favorable catalyst assessments associated with a lower probability of a downward revision. These patterns suggest that the relevance of each outlook dimension depends on the sector. This pattern can be understood in terms of differences in sector exposure. Materials and Industrials are particularly exposed to industry-wide demand, pricing, and cost conditions, making industry conditions (Q8) relevant to subsequent valuation updates. In Health Care and consumer sectors, company-specific developments such as clinical milestones, product launches, or changes in consumer demand may instead make catalysts (Q10) more informative about subsequent target-price actions.

\section{Applications}

\subsection{Multidimensional Analyst Outlook Indicators}

The question-based scores provide a multidimensional view of analyst judgments that a single target price cannot convey. Individual scores can serve as report-based indicators of analyst sentiment toward earnings, valuation, industry conditions, and other investment dimensions. The accompanying evidence allows users to inspect the textual basis of each assessment, supporting interpretation of the indicators.

\subsection{Industry and Theme Monitoring}

\begin{figure*}[t]
\centering
\includegraphics[width=\textwidth]{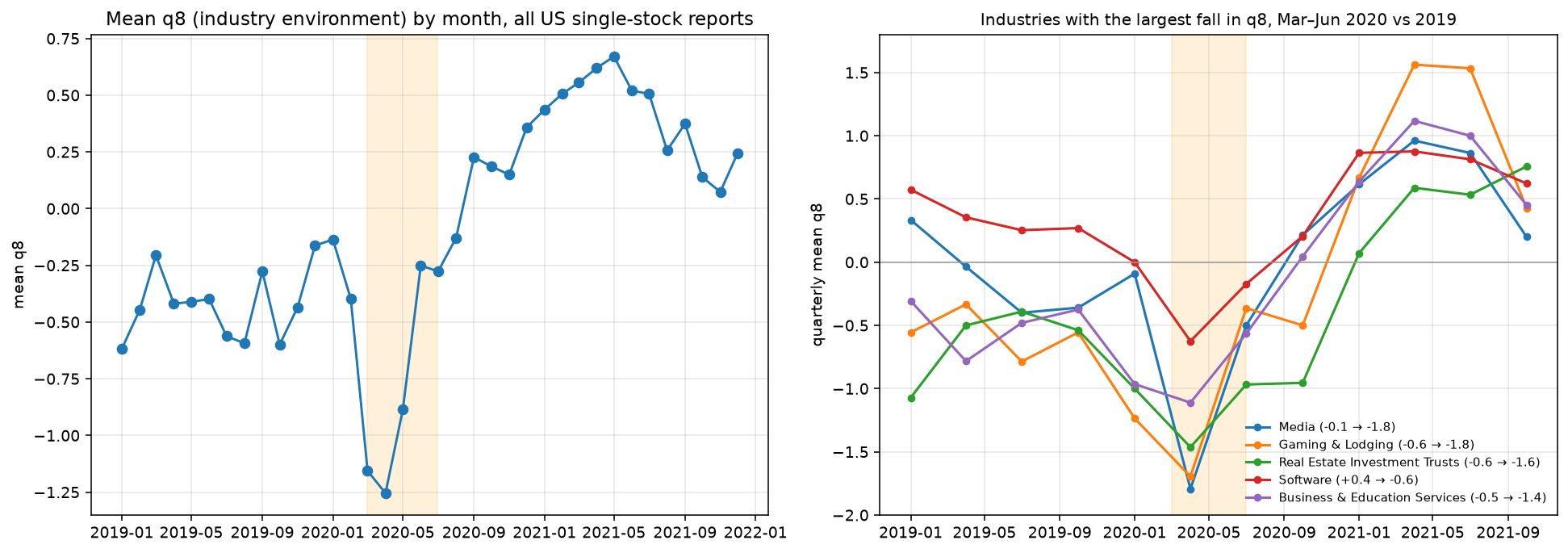}
\caption{Industry-environment scores (Q8) around the COVID-19 pandemic. Left: monthly mean scores across all reports in 2019--2021. Right: quarterly mean scores for the five industries with the largest decline from 2019 to March--June 2020, using non-null scores. Shading marks March--June 2020; legend values compare the 2019 and March--June 2020 means. Selected industries have at least 20 scored reports in 2019 and 15 in March--June 2020.}
\label{fig:q8-covid}
\Description{Monthly industry-environment scores in 2019 to 2021 and quarterly score trajectories for media, gaming and lodging, REITs, software, and business and education services, the five industries with the largest pandemic-period decline.}
\end{figure*}

Because the dimensions have fixed meanings, scores can be aggregated to examine how reported conditions vary across industries and over time. Figure~\ref{fig:q8-covid} illustrates this use with the industry-environment question (Q8) around the COVID-19 pandemic. The average monthly score falls from $-0.43$ in 2019 to $-1.25$ in April 2020 and rises to $+0.41$ in 2021. Comparing March--June 2020 with 2019, the mean score declines by 1.68 points for media, 1.19 for gaming and lodging, 1.01 for real estate investment trusts (REITs), and 0.89 for business and education services. These declines are consistent with pandemic-related pressures on advertising and live events, travel and leisure, commercial real estate, and in-person instruction, respectively.

\section{Discussion and Limitations}

We evaluate subsequent target-price actions rather than future stock returns because our objective is to capture the analyst's expressed outlook. Stock returns reflect many influences beyond that outlook, whereas subsequent target-price actions are more directly connected to the analyst's valuation judgments.

The study covers one brokerage and the US market. Analyses across additional brokerages and countries, together with validation of the supporting textual evidence, would help assess the generalizability of the findings. The extraction relies on a single LLM. Different LLMs may assign different scores or select different supporting evidence from the same report, potentially changing predictive performance and question-level interpretations. The robustness of these findings across LLMs remains to be evaluated.

Results may also depend on the choice and wording of questions. The ten questions cover investment dimensions that are broadly applicable across sectors. Sectors with distinct business models, such as real estate and banking, may require a sector-tailored approach that asks about their principal investment drivers.

\section{Conclusion}

We present a question-based representation of analyst outlooks for predicting subsequent target-price revisions. Target-price actions exhibit directional persistence, making the current revision a natural momentum baseline. In walk-forward evaluation, adding ten report-derived outlook scores improves prediction beyond current revision and sector controls, indicating that the narrative contains information about subsequent analyst actions beyond these observable characteristics. The scores outperform dictionary-based and FinBERT sentiment measures. Although high-dimensional embeddings achieve higher AUCs, the ten scores provide incremental predictive information in a compact representation.

Each score corresponds to an explicit investment question and is accompanied by supporting textual evidence. This structure allows users to examine which reported outlook dimensions are associated with subsequent revisions and inspect the basis of each assessment. The approach therefore combines predictive information with a directly interpretable representation of analyst sentiment, offering a useful complement to numerical target-price signals.

\bibliographystyle{ACM-Reference-Format}
\bibliography{draft}

\appendix

\section{Extraction Prompts}

\subsection{Questionnaire Prompt}

The questionnaire prompt specifies the scoring rules, response format, and ten investment questions. The report text replaces the content placeholder.

\begin{promptbox}{Scoring instructions}
\raggedright
\setlength{\parindent}{0pt}
\setlength{\parskip}{3pt}
Your task is to extract and quantify the views expressed by a sell-side equity research analyst in the provided research report. For each question, assess the direction and strength of the analyst's stated or clearly implied view regarding the company's financial performance, business outlook, valuation, and industry conditions. Base all assessments solely on evidence contained in the provided report. Do not form your own view of the company.\par
\smallskip
Rules:\par
1. Use only the provided report. Do not add outside knowledge or judgments not expressed in the report.\par
2. The company's name and ticker are masked as [COMPANY] and [TICKER]. Target-price figures, revision directions, and target-price-based percentages are masked as [TP], [TP\_ACTION], and [TP\_PCT]. Do not reconstruct or use the masked information as evidence.\par
3. Use a seven-point scale: -3 strongly negative, -2 moderately negative, -1 slightly negative, 0 neutral or unchanged, +1 slightly positive, +2 moderately positive, +3 strongly positive. Scores reflect the strength of the assessment expressed in the report, not your confidence in identifying it. Use 0 only for an explicitly stated or clearly implied neutral, unchanged, in-line, or balanced assessment. Use null for missing or unclear evidence; if the report does not discuss the topic, the score is null, never 0.\par
4. Evaluate each question independently using evidence that matches its comparison basis. For forward-looking questions, use the report's stated or clearly implied outlook and forecast periods.\par
5. For each question give the score and one line of evidence: a short quote or paraphrase of the passage in the report that supports the score. If the score is null, state briefly why (for example "not discussed"). Evidence of the form "not discussed" or "not mentioned" must be paired with a null score.\par
6. Return only the specified JSON object, complete and valid. Each value is an object with "score" (an integer from -3 to 3, or null) and "evidence" (one line of text; use single quotes if you quote the report). Do not include anything else.\par
\end{promptbox}

\begin{promptbox}{Response format}
\raggedright
\setlength{\parindent}{0pt}
\setlength{\parskip}{3pt}
\ttfamily\scriptsize
\{"q1": \{"score": null, "evidence": ""\},\par
"q2": \{"score": null, "evidence": ""\},\par
"q3": \{"score": null, "evidence": ""\},\par
"q4": \{"score": null, "evidence": ""\},\par
"q5": \{"score": null, "evidence": ""\},\par
"q6": \{"score": null, "evidence": ""\},\par
"q7": \{"score": null, "evidence": ""\},\par
"q8": \{"score": null, "evidence": ""\},\par
"q9": \{"score": null, "evidence": ""\},\par
"q10": \{"score": null, "evidence": ""\}\}\par
\end{promptbox}

\begin{promptbox}{Report input and questions}
\raggedright
\setlength{\parindent}{0pt}
\setlength{\parskip}{3pt}
\smallskip
Report:\par
\{\{content\}\}\par
\smallskip
Questions:\par
Q1 (EPS growth): Is the company's EPS expected to increase, remain unchanged, or decrease relative to the comparable prior-year period? Positive scores indicate growth; negative scores indicate decline.\par
\smallskip
Q2 (Revenue growth): Is the company's revenue expected to increase, remain unchanged, or decrease relative to the comparable prior-year period? Positive scores indicate growth; negative scores indicate decline.\par
\smallskip
Q3 (Profitability): Is the company's profitability, as measured by ROE, ROA, ROIC, or profit margins, expected to improve, remain unchanged, or deteriorate relative to the comparable prior-year period? Positive scores indicate improvement; negative scores indicate deterioration.\par
\smallskip
Q4 (Valuation): Does the report assess the company's current share price as undervalued, fairly valued, or overvalued? Positive scores indicate undervaluation; negative scores indicate overvaluation. Do not use masked target-price information or the investment rating alone as evidence.\par
\smallskip
Q5 (Earnings Consensus): Does the report expect the company's next reported EPS to be above, in line with, or below current market consensus? Positive scores indicate an expected beat; negative scores indicate an expected miss. Use null if no comparison with consensus is stated or clearly implied.\par
\smallskip
Q6 (Shareholder returns): Are dividends and share repurchases expected to increase, remain unchanged, or decrease relative to the comparable prior-year period? Positive scores indicate an increase; negative scores indicate a decrease.\par
\smallskip
Q7 (Competitive position): Is the company's market position or competitive advantage relative to its peers expected to strengthen, remain unchanged, or weaken? Positive scores indicate strengthening; negative scores indicate weakening.\par
\smallskip
Q8 (Industry environment): Are industry conditions facing the company expected to improve, remain unchanged, or deteriorate? Consider industry demand, pricing, competition, capacity, and input costs. Positive scores indicate more favorable conditions; negative scores indicate less favorable conditions.\par
\smallskip
Q9 (Government and regulatory environment): Are government policies or regulatory conditions facing the company expected to become more favorable, remain unchanged, or become less favorable? Positive scores indicate more favorable conditions; negative scores indicate less favorable conditions.\par
\smallskip
Q10 (Catalysts): Does the report identify specific future events or developments as catalysts for the shares? Score their expected overall share-price impact: positive scores indicate a positive impact; negative scores indicate a negative impact; 0 indicates an explicitly neutral or balanced impact. The timing of a catalyst does not need to be known. Use null if no catalyst is discussed or the expected impact is unclear.\par
\end{promptbox}

\subsection{Masking Instructions}

The masking prompt identifies spans for replacement while preserving the surrounding text. The covered-company identifier and text-block identifiers are supplied with the report. The instructions are reproduced below, followed by an example of the output format.

\begin{promptbox}{Masking instructions}
\raggedright
\setlength{\parindent}{0pt}
\setlength{\parskip}{3pt}
Your task is to identify exact text spans to mask in equity research report blocks. Return masking instructions only; do not rewrite the report.\par
\smallskip
Covered company: \{\{company\}\}\par
\smallskip
Mask:\par
* The covered company's name and aliases as [COMPANY].\par
* The covered company's ticker and security identifiers as [TICKER].\par
* Current, previous, and bull/base/bear target-price amounts, including explicit target-price change amounts, as [TP].\par
* Expressions explicitly indicating a target-price increase, decrease, or maintenance as [TP\_ACTION].\par
* Target-price percentage changes and target-price-based upside or downside, including their directional labels, as [TP\_PCT].\par
\smallskip
Rules:\par
1. Mask by context, not numeric identity. Preserve non-target-price financial figures, assumptions, and revisions, as well as other companies' identifiers.\par
2. Select the smallest complete masking span and copy it exactly, preserving case, punctuation, and whitespace. Do not reconstruct missing or corrupted text.\par
3. For each span, return block\_id, text, occurrence, and replacement. Use only supplied block IDs. occurrence is the 1-based index among all exact, case-sensitive, non-overlapping matches of text in the original block, including unmasked matches.\par
4. Return one object per selected occurrence, with no duplicates or overlapping spans.\par
5. Return only JSON in the format below. If no masking is needed, return \{"masks": []\}.\par
\end{promptbox}

\begin{promptbox}{Masking response example}
\raggedright
\setlength{\parindent}{0pt}
\setlength{\parskip}{3pt}
\ttfamily\scriptsize
\{\par
"masks": [\par
\{\par
"block\_id": "summary:1:1",\par
"text": "\$119",\par
"occurrence": 1,\par
"replacement": "[TP]"\par
\}\par
]\par
\}\par
\end{promptbox}

Each instruction is checked for an exact match to the specified occurrence in the original block. The replacements change only the selected spans; the remaining text is retained for scoring.

\begin{figure*}[!t]
\centering
\includegraphics[width=\textwidth]{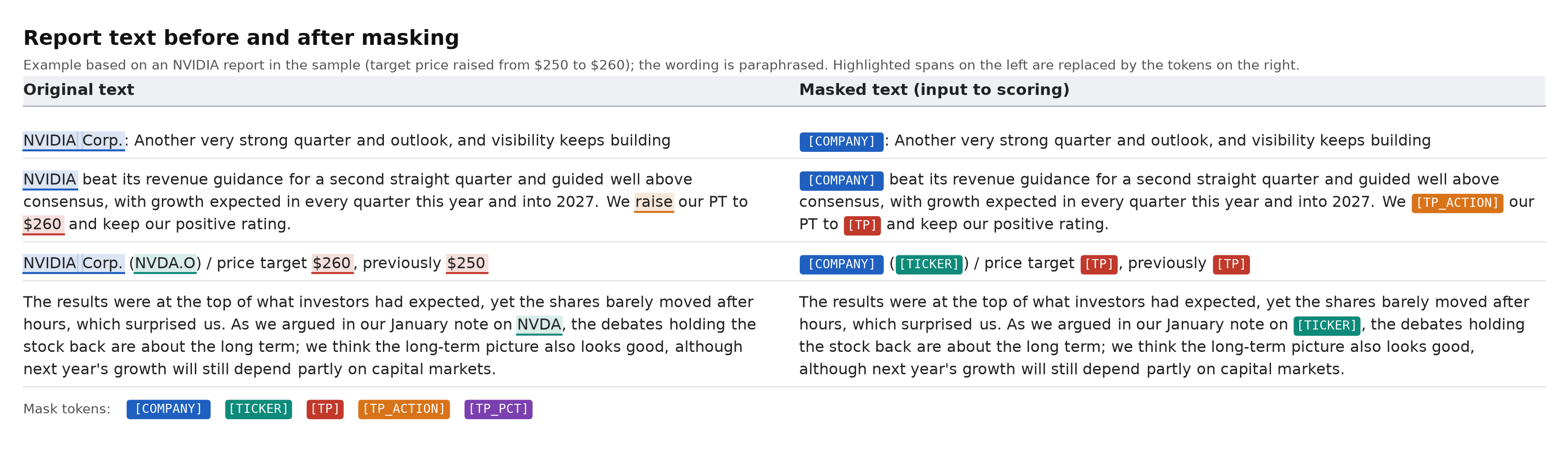}
\caption{Illustrative passages based on an NVIDIA report before and after masking. Wording is paraphrased for presentation; colors identify masking categories.}
\label{fig:masking-example}
\Description{Paraphrased illustrative passages before and after masking}
\par\medskip
\includegraphics[width=\textwidth]{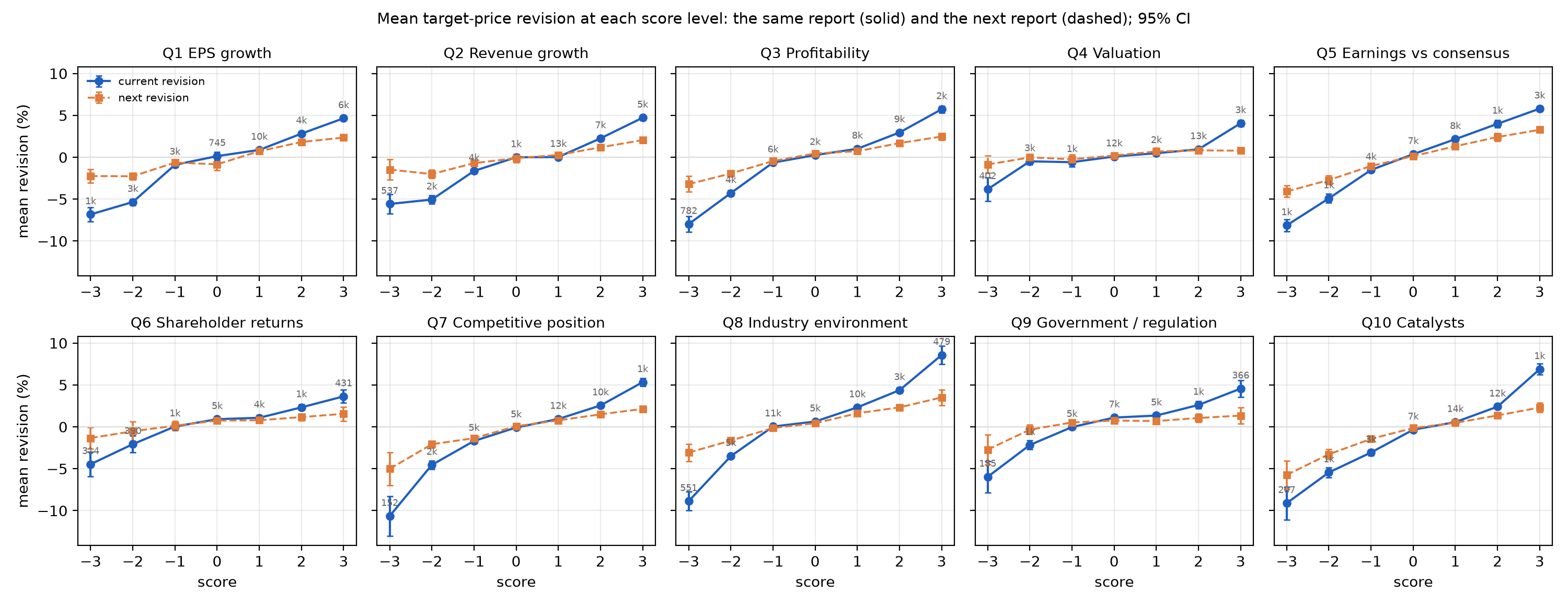}
\caption{Mean target-price revision (\%) by question score in the 2017--2026 sample. Blue solid lines show contemporaneous revisions; orange dashed lines show subsequent revisions. Error bars indicate 95\% confidence intervals. Current-revision means use the final sample of 41,640 reports; subsequent-revision means additionally exclude absolute subsequent revisions of 50\% or more. Labels show current-revision observation counts at each score level; null scores are not plotted.}
\label{fig:score-current-revision}
\Description{Ten panels compare mean contemporaneous revisions in solid blue with subsequent revisions in dashed orange at score levels from minus three to plus three, with confidence intervals.}
\end{figure*}

\end{document}